\documentclass[aps,prd,twocolumn,superscriptaddress]{revtex4-2}
\usepackage{graphicx}
\usepackage{amsmath,amssymb,amsfonts}
\usepackage{comment}
\usepackage{float}
\usepackage{placeins}
\usepackage{braket}
\usepackage{array}
\usepackage[colorlinks,
	linkcolor=blue,
	citecolor=blue,
	urlcolor=blue]{hyperref}
\usepackage{url}
\usepackage{cleveref}

\begin{document}

\title{Accurate calibration method of the Back-linked Fabry Perot interferometer using null channel for space-based gravitational wave antenna}

\author{Yusuke Okuma}
\email[]{okuma-yusuke319@g.ecc.u-tokyo.ac.jp}
\affiliation{Department of Physics, The University of Tokyo, Bunkyo, Tokyo 113-0033, Japan}

\author{Shoki Iwaguchi}
\affiliation{Department of Space Astronomy and Astrophysics, Institute of Space and Astronautical Science, JAXA }

\author{Ryosuke Sugimoto}
\affiliation{Department of Physics, The University of Tokyo, Bunkyo, Tokyo 113-0033, Japan}

\author{Yutaro Enomoto}
\affiliation{Department of Physics, The University of Tokyo, Bunkyo, Tokyo 113-0033, Japan}
\affiliation{Department of Space Astronomy and Astrophysics, Institute of Space and Astronautical Science, JAXA }

\author{Kiwamu Izumi}
\affiliation{Department of Physics, The University of Tokyo, Bunkyo, Tokyo 113-0033, Japan}
\affiliation{Department of Space Astronomy and Astrophysics, Institute of Space and Astronautical Science, JAXA }

\date{\today}

\begin{abstract}
\noindent Accurate calibration of the detector response is essential for precision gravitational-wave (GW) observations. In the Back-linked Fabry-Perot Interferometer (BLFPI), which employs Fabry-Perot cavities to achieve high GW sensitivity, accurate calibration of the cavity response is particularly important because residual calibration errors directly limit the suppression of laser frequency noise. However, no practical calibration strategy for the cavity response has yet been established. In this paper, we propose a method for calibrating the cavity response using a null channel constructed from the three interferometer outputs of the triangular laser-link configuration. In null channel, GW signals are canceled and residual laser frequency noise indicates errors in the cavity response estimates. Therefore we can estimate the cavity response by optimizing the response parameters to minimize the integrated power of the null channel. We derive the estimation bias introduced by technical noise analytically and evaluate the performance of the proposed method numerically. The results show that the residual laser frequency noise after calibration is more than an order of magnitude below the designed BLFPI sensitivity throughout the observation band. This demonstrates that the proposed null-channel calibration method achieves the accuracy required for effective subtraction of laser frequency noise.

\end{abstract}

\maketitle

\section{Introduction}
Since the first detection of gravitational waves in 2015~\cite{abbott2016observation}, gravitational-wave~(GW) astronomy has rapidly advanced. The network of ground-based observatories, including LIGO~\cite{ligo2015advanced}, Virgo~\cite{acernese2015advanced}, and KAGRA~\cite{aso2013interferometer}, have detected hundreds of GW events~\cite{abac2025gwtc} in the frequency band around 100~Hz. At nanohertz frequencies, Pulsar Timing Array collaborations have reported evidence for a stochastic GW background~\cite{antoniadis2023second,agazie2023nanograv,miles2025meerkat,reardon2023search,xu2023searching}. To bridge the frequency gap between ground-based detectors and PTAs, several space-based GW observatories have been proposed. While proposed observatories such as LISA~\cite{amaro2017laser}, TianQin~\cite{luo2016tianqin} and Taiji~\cite{hu2017taiji} employ optical transponders for inter-satellite interferometry, other missions including DECIGO and B-DECIGO~\cite{kawamura2011japanese, komori2026current} utilize Fabry-Perot cavities to enhance strain sensitivity to gravitational waves in the decihertz frequency band in a way similar to those used in ground-based observatories.

To improve the feasibility of space-based Fabry-Perot GW observatories, an interferometer topology, the Back-linked Fabry-Perot Interferometer (BLFPI), has been proposed~\cite{izumi2021back}. The BLFPI employs two independent lasers and forms two Fabry--Perot cavities in its arms. Because the BLFPI maintains the cavity resonance by controlling the laser frequencies, the cavity length control via the actuation of the position of the test mass is not required. The differential phase fluctuations of the two lasers are measured by taking a beat note of the laser fields. This topology provides both high GW sensitivity through the use of Fabry-Perot cavities and relaxed control requirements.

For the BLFPI, as well as other interferometers such as Time-Delay Interferometry (TDI) employed for LISA~\cite{tinto2014time} and the Differential Fabry--Perot Interferometer (DFPI), which is another candidate topology for DECIGO~\cite{nagano2021demonstration}, suppression of laser frequency noise is a major challenge. In interferometers employing Fabry--Perot arms, such as the DFPI and the BLFPI, the laser frequency control to the arm cavity resonance provides partial suppression of laser frequency noise. However, relying solely on laser frequency control to achieve the required frequency-noise suppression in DECIGO and B-DECIGO would require extremely high control-loop gains of approximately $1.7\times 10^9$ and $1.7\times 10^8$, respectively~\cite{michimura2025initial}. Furthermore, the low cavity-pole frequencies associated with the long arm cavities e.g., 100~km or 1000~km would constrain the achievable feedback bandwidth and stability~\cite{Nagano2020}. As a complementary approach to mitigate the impact of residual laser frequency noise, in the TDI, the DFPI, and the BLFPI, the transfer functions relating the laser frequency noise to the interferometer outputs are estimated. Subsequently the noise is subtracted by appropriately combining multiple signals.

For the BLFPI, since it utilizes two independent lasers, common-mode rejection of laser frequency noise is not applicable unlike for DFPI. To suppress the laser frequency noise, the BLFPI combines the beat signal with two Pound--Drever--Hall (PDH) signals~\cite{drever1983laser} of the arm cavities. Then, we have to estimate the cavity response, which is more involved than the time-delay transfer functions estimated in the TDI.

The ability to subtract the frequency noise has been experimentally demonstrated using one BLFPI, achieving approximately 46~dB of frequency-noise suppression~\cite{sugimoto2024experimental}. However, in this demonstration, the response-estimation procedure did not account for the challenges mentioned in the next paragraph. To date, a practical strategy for estimating these responses has not yet been established.

Two major challenges arise when we estimate the cavity responses in the BLFPI. The first is the presence of GW signals, and the second is the limited hardware resources available aboard the spacecraft for calibration. At the frequencies below approximately 10~Hz, where the BLFPI is designed to achieve high sensitivity, GW signals are expected to be continuously present in the observational band~\cite{kawamura2021current}. Conventional calibration methods typically compare injected calibration signals with detector outputs to derive the transfer function of a cavity. Therefore, the presence of GW signals can contaminate the calibration signals and degrade calibration accuracy.

The second challenge is the substantial hardware required for conventional calibration techniques. Ground-based detectors mainly employ photon calibrators and Newtonian calibrators to characterize cavity responses~\cite{karki2016advanced,ross2021initial}. These techniques induce known displacements of the cavity mirrors and compare them against the detector output to estimate the detector's response. Although they have achieved remarkable calibration accuracy~\cite{sun2021characterization,aubin2024virgo}, the required calibration hardware may be too large and complex to be accommodated aboard a spacecraft.

To address these challenges, we investigate the use of the null channel for calibrating the cavity response. The null channel is constructed by combining signals from three interferometers such that GW signals cancel out. Null channels have been proposed for use in the Einstein Telescope (ET)~\cite{goncharov_utilizing_2022,regimbau_mock_2012,narola_null-stream-based_2025}, LISA~\cite{muratore_effectiveness_2023}, and networks of GW detectors~\cite{schutz_self-calibration_2020,gursel_near_1989}. Previous studies have demonstrated their usefulness for noise characterization, detector calibration, and parameter estimation of GW sources.

In this paper, we propose a method for calibrating the cavity response in the BLFPI utilizing the null channel. The baseline BLFPI mission concept consists of three interferometers, enabling the construction of the null channel in which GW signals are canceled. Furthermore, we propose utilizing the laser frequency noise in the BLFPI as a calibration signal as opposed to adding new hardware. Our method constructs the null channel and evaluates its integrated power (IP). If the assumed cavity responses differ from the true ones, residual frequency noise remains in the null channel and elevates the IP. As the estimated cavity responses approach the true ones, the residual frequency noise is reduced, leading to a smaller IP of the null channel without the contamination from GW signals. In this way, the cavity responses can be calibrated by optimizing the response parameters to minimize the IP of the null channel. We demonstrate the validity of this working principle through numerical simulation.

This paper is organized as follows. In Sec.~\ref{sec:BLFPI}, we introduce the BLFPI topology and review the post-processing technique used for frequency-noise suppression. In Sec.~\ref{sec:Null}, we present our calibration method utilizing the null channel. In Sec.~\ref{sec:num}, we evaluate its performance through numerical simulations. In Sec.~\ref{sec:discussion}, we discuss the developments required to apply this method to future space-based GW detectors. Finally, Sec.~\ref{sec:conclusion} concludes our study.

\section{Postprocessing in the Back-linked Fabry-Perot Interferometer\label{sec:BLFPI}}
\begin{figure}[t]
    \centering
    \includegraphics[width=0.7\linewidth]{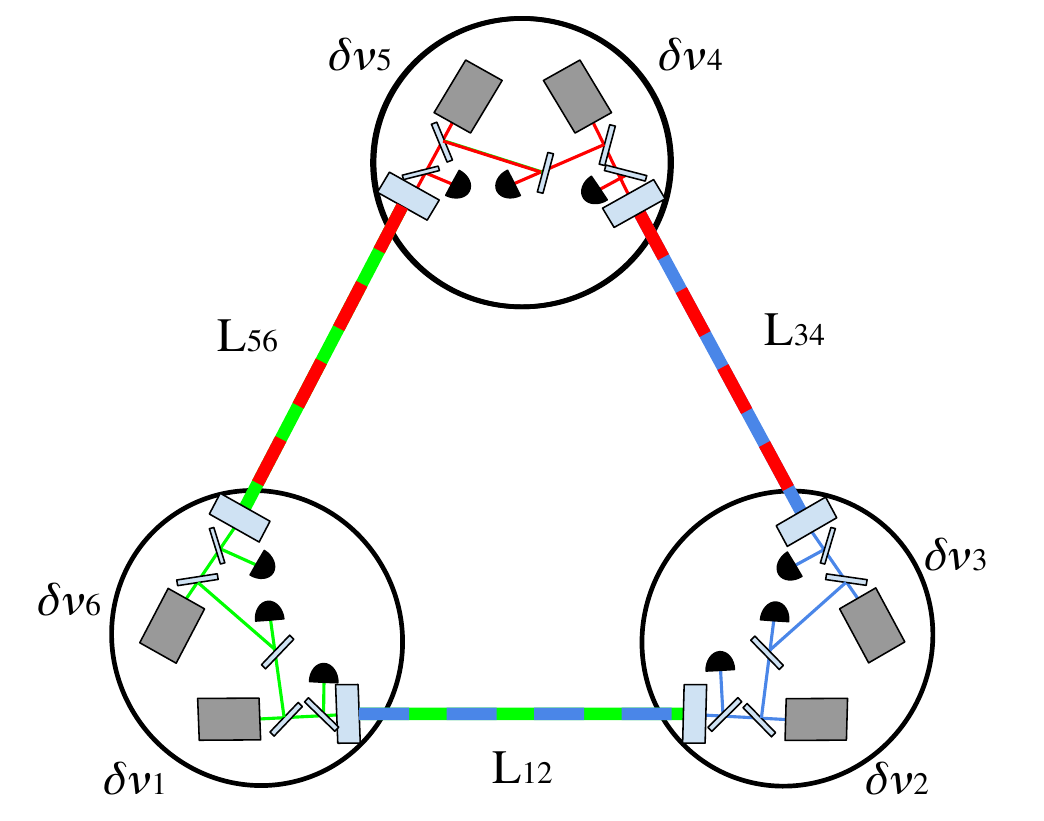}
    \caption{Conceptual schematic of the BLFPI}
    \label{fig:iir:sc_null}
\end{figure}
The BLFPI is an interferometer topology that employs two independent lasers and forms two Fabry--Perot cavities, one in each arm. Figure~\ref{fig:iir:sc_null} illustrates the conceptual design of the BLFPI. A single interferometer consists of two lasers and two cavities. Each laser is frequency-locked to its corresponding cavity. The cavity resonance is maintained by controlling the laser frequency. As a result, precise active control of the cavity lengths by actuating the masses is no longer required, which is one of the most attractive features of this topology for implementing the inter-satellite Fabry-Perot cavities.

Gravitational waves induce fluctuations in the resonant frequencies of the arm cavities. By measuring the beat frequency between the two lasers propagating in different arms, these GW signals can be extracted. As is the case for LISA and the DFPI, the BLFPI is also affected by laser frequency noise that contaminates the GW signal. To suppress this noise, the BLFPI combines three signals: the beat signal and two PDH signals of the arm cavities. The beat signal contains the frequency noise of the two lasers together with the GW signal. The PDH signals also contain laser frequency noise and GW signals, but with different relative contributions compared with the beat signal. Therefore, the laser frequency noise can be subtracted from the beat signal using the two PDH signals after correcting them by applying the inverse of the cavity responses.

\begin{figure*}[!t]
        \centering
    \includegraphics[width=1\linewidth]{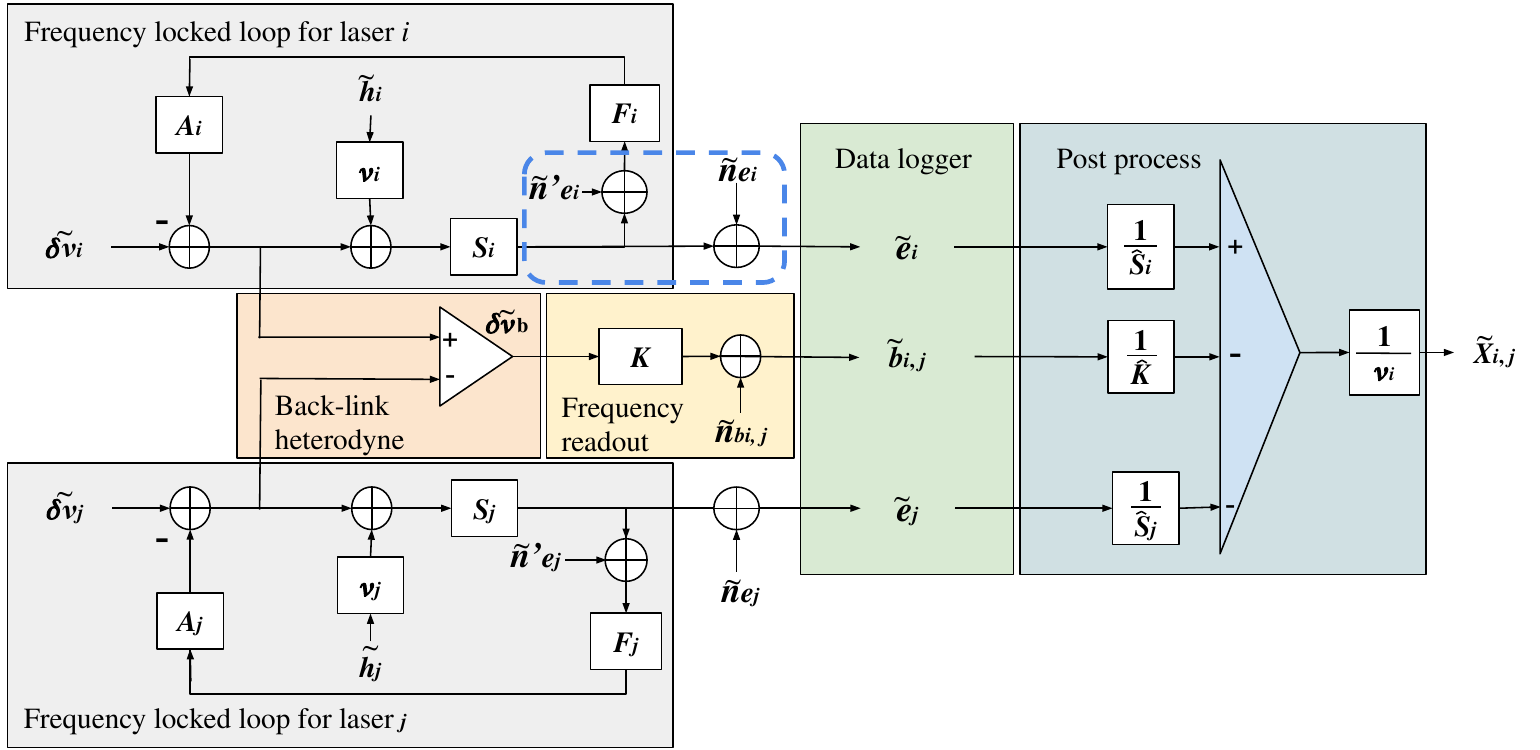}
    \caption{Block diagram of the entire system of one BLFPI. The frequency-locking loop for laser $i$ consists of cavity response $S_i$, filter $F_i$ and laser frequency actuator $A_i$. Two control loops maintain the resonance of the two lasers. Back-link heterodyne system reads out the beat frequency of the resonant laser fields with the response of $K$. We apply the inverse responses of the estimated response of the frequency readout system $\hat{K}$ to the beat signal and the inverse of the estimated response of the cavities $\hat{S}_i$ to each PDH signal, respectively. The corrected signals are then combined and converted into strain by applying the inverse of the laser frequency. Then, we obtain the gravitational wave signals.}
    \label{fig:block}
\end{figure*}
Figure~\ref{fig:block} shows a block diagram of the signal processing scheme of the BLFPI. In the post-processing stage, the inverse of the estimated frequency-readout response $\hat{K}$ and the inverse of the estimated cavity responses $\hat{S}_i$ are applied to the beat signal and the two PDH signals, respectively. These corrected signals are then combined and converted into strain units to obtain the reconstructed GW signal $\Tilde{X}_{i,j}$. Quantities with a tilde denote frequency-domain variables throughout this paper. For a BLFPI consisting of lasers $i$ and $j$, the reconstructed signal can be written as

\begin{eqnarray}
    &&\Tilde{X}_{i,j}\notag\\
    &=&\frac{\Tilde{b}_{i,j}}{\nu_i \hat{K}}-\frac{\Tilde{e}_i}{\nu_i \hat{S}_i}+\frac{\Tilde{e}_{j}}{\nu_i \hat{S}_j}\\
    &=&\frac{1}{\nu_i}\sum_{i,j}(-1)^i\left\{\frac{G_i}{(1+G_i)}\left(1-\frac{S_i}{\hat{S}_i}\right)\left(\frac{\Tilde{\delta\nu}_i}{G_i}-\frac{\Tilde{n'}_{\mathrm{e}i}}{S_i}\right)\right.\notag\\
    &&\left.-\Tilde{h}_i+\frac{\Tilde{h}_i}{1+G_i}\left(1-\frac{S_i}{\hat{S}_i}\right)-\frac{\Tilde{n}_{\mathrm{e}i}}{\hat{S}_i}+\Tilde{n}_{\mathrm{b}ij}\right\}\label{eq:iir:residual},
\end{eqnarray}
where ($G_i \equiv A_i F_i S_i$) is the open-loop transfer function. The quantities $\Tilde{e}_i$, $\Tilde{b}_{i,j}$, $\Tilde{h}_i$, $\Tilde{\delta\nu}_i$, $\Tilde{n'}_{\mathrm{e}i}$, $\Tilde{n}_{\mathrm{e}i}$, $\Tilde{n}_{\mathrm{b}ij}$, and $\nu_i$ denote the PDH signal, the beat signal between lasers $i$ and $j$, the GW signal, the laser frequency noise, the in-loop error noise, the out-of-loop error noise, the beat noise, and the laser frequency, respectively. We assume that the difference between the estimated response of the frequency-readout system $\hat{K}$ and the true response $K$ is small enough and the difference of the frequency of two lasers $\nu_i$ and $\nu_j$ is controlled to be smaller than approximately 100~MHz so that the beat signal remains within the high-sensitivity bandwidth of the photo detectors. Therefore, comparing the difference frequency to the lasers' frequency about 600~THz, we assume $\nu_i=\nu_j$. The first term in curly bracket in Eq.~(\ref{eq:iir:residual}) represents the residual laser frequency noise and in-loop error noise after the post process, while the second represents the GW signal and the third the calibration error in the reconstructed GW signals introduced by imperfect response estimation. If the estimated cavity response was identical to the true one, the frequency noise and the in-loop error noise would be completely canceled. However, any discrepancy between the estimated and true responses results in residual noise. Since the in-loop error noise $\tilde n'_{ei}/S_i$ is designed to be more than one order of magnitude smaller than the suppressed laser frequency noise $\tilde{\delta\nu_i}/G_i$, we neglect it hereafter.

Assuming that the estimated cavity response contains a relative magnitude error $\delta A_i$ and a phase error $\delta\phi_i$, one can express the residual frequency-noise ratio as

\begin{eqnarray}
    \left|1-\frac{S_i}{\hat{S}_i}\right|&=&\left|1-\frac{1}{(1+\delta A_i)e^{i\delta \phi_i}}\right|
    \simeq \sqrt{\delta A_i^2+\delta\phi_i^2}\label{eq:iir:residual_ratio}
\end{eqnarray}
Equation~(\ref{eq:iir:residual_ratio}) indicates that both the magnitude error $\delta A_i$ and the phase error $\delta\phi_i$ must be sufficiently small to achieve a high level of frequency-noise suppression. When $\delta A_i \simeq \delta\phi_i$, achieving the suppression factor for frequency noise $R_{\mathrm{sub}}\equiv 1/\left|1-\frac{S_i}{\hat{S}_i}\right|$ requires

\begin{eqnarray}
    \delta A_i\sim \delta \phi_i<\frac{1}{\sqrt{2}R_{\mathrm{sub}}}.
\end{eqnarray}
$R_{\mathrm{sub}}$ is sought to be a factor of $10^4$ in the proposal paper~\cite{izumi2021back} assuming a laser frequency noise amplitude spectral density of approximately $0.1~\mathrm{Hz/\sqrt{Hz}}$ at 1~Hz, which is available at laboratory experiment~\cite{numata2004thermal}. Although such accurate estimation is required, a practical calibration strategy has not yet been established.

It is worth noting that the accuracy required for cavity-response estimation differs depending on its purpose. For the reconstruction of GW signals, as shown in the third term in Eq.~(\ref{eq:iir:residual}), cavity-response estimation errors have a much smaller impact. Because the magnitude of the open-loop gain $|G|$ is typically very large in the frequency range of interest, e.g., $|G| \sim 10^4$, the required accuracy of the cavity-response estimation for signal reconstruction is significantly less stringent than those for frequency-noise suppression.

\section{Estimation of the cavity response using Null channel\label{sec:Null}}
As discussed in the previous section, accurate estimation of the cavity response is essential for effective frequency-noise subtraction. At the same time, the presence of GW signals complicates the calibration process. In this section, we introduce a method utilizing null channel to estimate the cavity response while suppressing the influence of GW signals.

After the post process, each of three interferometer produces a GW data stream with reduced frequency noise. By summing these three data streams, the null channel can be constructed in which the GW signals vanish. The null channel $V_{\mathrm{null}}$ is given by

\begin{eqnarray}
    &&\Tilde{V}_\mathrm{null}\notag\\
    &=& \Tilde{X}_{6,1}+\Tilde{X}_{2, 3}+\Tilde{X}_{4,5}\\
    &=&\frac{1}{\nu}\left\{\sum_{i=1}^6(-1)^{i+1}\frac{\Tilde{\delta\nu}_i}{(1+G_i)}\left(1-\frac{S_i}{\hat{S}_i}\right)\right.\notag\\
    &&-\sum_{i=1}^6(-1)^{i}\left(1-\frac{S_i}{\hat{S}_i}\right)\frac{\Tilde{h}_i\nu_i}{(1+G_i)}\notag\\
    &&-\sum_{i=1, 3,5}(-1)^{i}(\Tilde{h}_i\nu_i-\Tilde{h}_{i+1}\nu_{i+1})\notag\\
    &&\left.+\sum^6_{i=1}(-1)^i\frac{\Tilde{n}_{\mathrm{e}i}}{\hat{S}_i}+\sum_{i, j}\Tilde{n}_{\mathrm{b}ij}\right\}.\label{eq:null:null}
\end{eqnarray}
We again assumed $\nu_i\simeq\nu$.

Because the laser frequency noise is expected to be much larger than the GW signal, the contribution of the second term in Eq.~(\ref{eq:null:null}) is negligible compared with the first term for the purpose of calibration. Furthermore, if the two lasers injected from the opposite side of the same cavity experienced nearly identical GW-induced phase perturbations, the third term would also negligible.

Strictly speaking, the third term of Eq.~(\ref{eq:null:null}) does not vanish completely. Even within the same cavity, photons propagating in opposite directions experience slightly different phase perturbations by gravitational waves. This effect becomes more significant at higher frequencies, particularly near and above the cavity free spectral range (FSR)~\cite{rakhmanov2005response}. Similar limitations of utilizing null channels have been discussed for the Einstein Telescope~\cite{virtuoso_beyond_2025}.

For the BLFPI with the cavity length of 100~km, the cavity FSR is approximately 1.5~kHz, which is more than two orders of magnitude higher than the observation band below 10 Hz. At frequencies more than two orders of magnitude below the FSR, the integrated power of the null channel relative to that of the GW signals is about $10^{-4}$. Thus, the residual GW signal in the null channel is negligible in practice~\cite{virtuoso_beyond_2025}. Therefore, the third term of Eq.~(\ref{eq:null:null}) is neglected in the reminder of this paper.

We estimate the cavity responses by minimizing the IP of the null channel,

\begin{eqnarray}
    &&\mathrm{IP}(\Tilde{V}_\mathrm{null})\notag\\
    &=&\frac{1}{2\pi\nu^2}\left[\int_{0}^\infty d\omega \sum^6_{i=1}\left|\left(1-\frac{S_i}{\hat{S}_i}\right)\frac{\Tilde{\delta\nu}_i}{(1+G_i)}\right|^2\right.\notag\\
    &&\left.+\sum_{i,j}\left|\Tilde{n}_{\mathrm{b}ij}\right|^2+\sum_{i=1}^6\left|\frac{\Tilde{n}_{\mathrm{e}i}}{\hat{S}_i}\right|^2\right]\label{eq:iir:null}
\end{eqnarray}
We assume that the technical noises and laser frequency noises are mutually uncorrelated. The first term in Eq.~(\ref{eq:iir:null}) vanishes only when the estimated cavity responses are identical to the true one. Therefore our idea is that the cavity responses can be estimated by minimizing $\mathrm{IP}(\Tilde{V}_{\mathrm{null}})$.

In practice, however, the third term in Eq.~(\ref{eq:iir:null}) also depends on the estimated cavity response. Therefore, the IP is not necessarily minimized when the estimated response exactly matches the true response. Technical noise consequently introduces a bias in the estimated cavity response. Below, we analytically evaluate such a bias and the resulting residual noise.

To obtain analytical insight into the estimation bias, we here consider each frequency component independently. Note that the actual estimation simultaneously uses all frequency components. For a laser $i$ at a given frequency, the IP contribution can be written using the magnitude error $\delta A_i$ and the phase error $\delta \phi_i$ of the cavity response function as

\begin{eqnarray}
    &&\left|1-\frac{e^{-i\delta \phi_i}}{1+\delta A_i}\right|^2\frac{|\Tilde{\delta\nu}_i|^2}{|1+G_i|^2\nu^2}+\left|\frac{e^{-i\delta \phi_i}}{1+\delta A_i}\frac{\Tilde{n_{\mathrm{e}i}}}{S_i\nu}\right|^2\notag\notag\\
    &&+\frac{|\Tilde{n_{\mathrm{b}ij}}|^2}{2}\label{eq:iir:one_res}.
\end{eqnarray}
We assume that each laser contributes half of the beat-noise power. Defining

\begin{eqnarray}
    g=\frac{e^{-i\delta \phi_i}}{1+\delta A_i}
\end{eqnarray}
and differentiating Eq.~(\ref{eq:iir:one_res}) with respect to $g$, we find that Eq.~(\ref{eq:iir:one_res}) attains its minimum at

\begin{eqnarray}
    g=\frac{\frac{\left|\Tilde{\delta\nu}_i\right|^2}{|1+G_i|^2}}{\frac{|\Tilde{\delta\nu}_i|^2}{|1+G_i|^2}+\left|\frac{\Tilde{n}_{\mathrm{e}i}}{S_i}\right|^2}.
\end{eqnarray}
This yields the estimation bias

\begin{eqnarray}
    \hat{\delta A_i}&=&\frac{\left|\frac{\Tilde{n_{\mathrm{e}i}}}{S_i}\right|^2}{\frac{|\Tilde{\delta\nu}_i|^2}{|1+G_i|^2}}\label{eq:iir:biased}
\end{eqnarray}
and

\begin{eqnarray}
    \hat{\delta\phi_i}=0.
    \label{eq:iir:biasp}
\end{eqnarray}
These analysis clarified that the estimation of the cavity response utilizing the null channel would result in the biased estimation.

Substituting these biased estimates into Eq.~(\ref{eq:iir:one_res}), the IP contribution becomes

\begin{eqnarray}
    \left|\frac{\Tilde{n_{\mathrm{e}i}}}{S_i}\right|^2\frac{1}{1+\frac{\left|\frac{\Tilde{n_{\mathrm{e}i}}}{S_i}\right|^2}{\frac{|\Tilde{\delta\nu}_i|^2}{|1+G_i|^2}}}+\frac{|\Tilde{n_{\mathrm{b}ij}}|^2}{2}.\label{eq:null:noise_opt}
\end{eqnarray}
For a BLFPI which consists of lasers $i$ and $j$, the noise floor is the sum of Eq.~(\ref{eq:null:noise_opt}) for $i$ and $j$.
\begin{eqnarray}
    \sum_{i,j}\left\{\left|\frac{\Tilde{n_{\mathrm{e}i}}}{S_i}\right|^2\frac{1}{1+\frac{\left|\frac{\Tilde{n_{\mathrm{e}i}}}{S_i}\right|^2}{\frac{|\Tilde{\delta\nu}_i|^2}{|1+G_i|^2}}}+\frac{|\Tilde{n_{\mathrm{b}ij}}|^2}{2}.\right\}\label{eq:iir:opt_noise}
\end{eqnarray}
This result implies that the noise floor of the BLFPI can be reduced by appropriately tuning $\hat{S}_i$. At first glance, this result appears counterintuitive because it seemingly indicates that the noise floor dominated by shot noise can be reduced by biased estimation of the cavity response. This apparent reduction does not arise from a decrease in the shot noise itself. Instead, it originates from the design of the control system where we divide the reflected laser from the cavity at a beam splitter and measure the PDH signal at two different photo detectors (PD) as shown in Fig.~\ref{fig:block}. Eq.~(\ref{eq:iir:opt_noise}) indicates that the signal-to-shot-noise ratio at the out-of-loop PD that is once degraded by beam splitting can be partially recovered by the optimal choice of $\hat{S}_i$. The detailed discussion is given in Appendix~\ref{sec:app:noise}.

The above analysis treated each frequency independently. In practice, however, it is effective to estimate the parameterized response function by synthesizing the information of all the frequency components. Consequently, the analytically derived bias does not necessarily appear in the full estimation procedure. Such a bias appears only if the parameter mismatch produces frequency-dependent deviations consistent with Eqs.~(\ref{eq:iir:biased}) and (\ref{eq:iir:biasp}). The validity of this single-frequency approximation is examined through numerical simulations in Sec.~\ref{sec:num}.

\section{Numerical experiment\label{sec:num}}
\begin{figure*}[!t]
    \centering
    \includegraphics[width=1\linewidth]{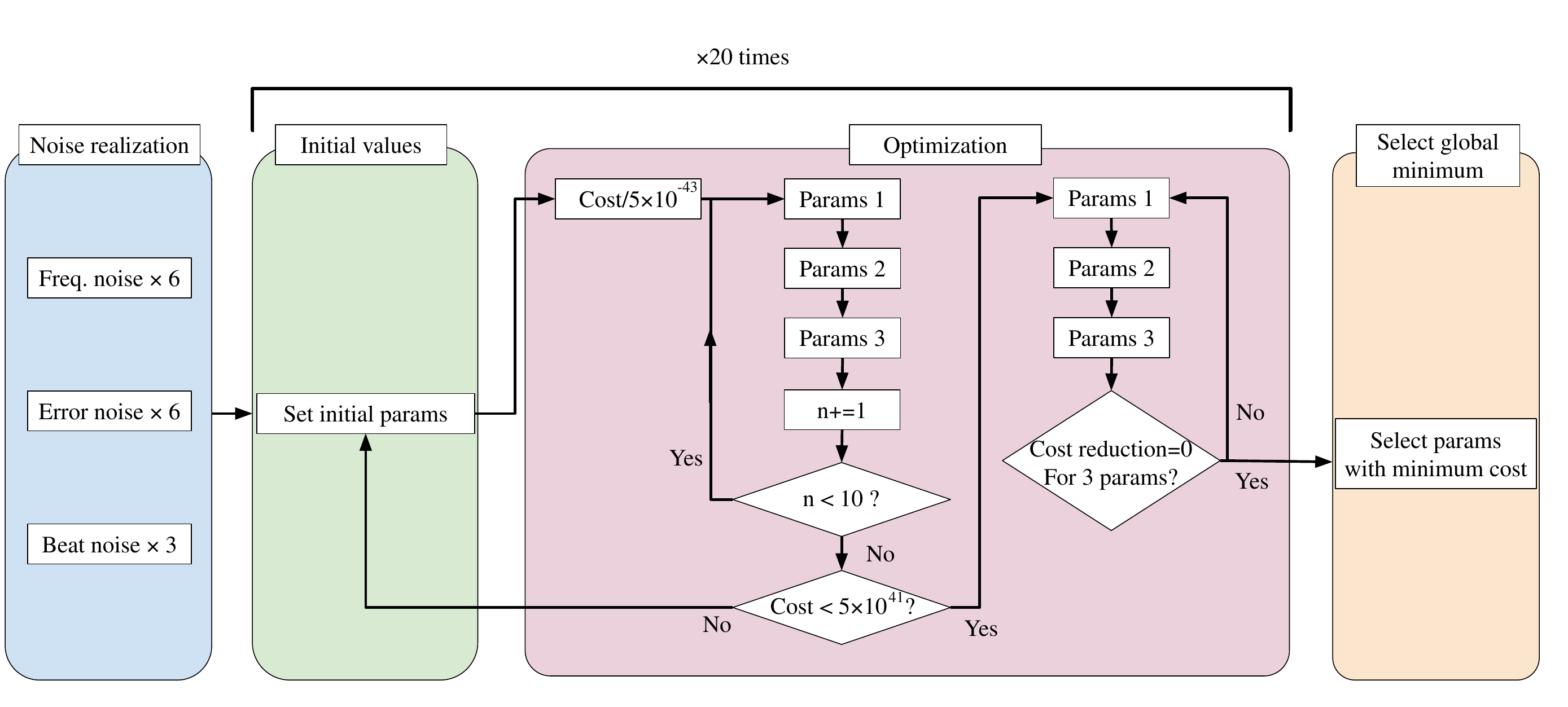}
    \caption{Flowchart of the optimization strategy for $\hat{S}_i$. First, six realizations of the laser frequency noise, six realizations of the PDH error noise, and three realizations of the beat noise are generated. Twenty independent initial parameter sets are then generated with random variation. The cost function is evaluated and minimized by optimizing the parameters. In the optimization of the parameters, we classified the parameters into three groups and optimized each group one by one. Parameter group $i$ consists of the parameters associated with cavity $i$ and the lasers which are injected into cavity $i$. Finally, the solution with the lowest cost is selected.}
    \label{fig:strategy}
\end{figure*}

\begin{table}[b]
    \centering
    \begin{tabular}{c|c|c|c}
        \hline
         & Cavity 1 & Cavity 2 & Cavity 3 \\
         & (Params 1) & (Params 2) & (Params 3)\\
        \hline
        $r_\mathrm{i}$ & 0.98 & 0.984 & 0.99\\
        $r_\mathrm{e}$ & 0.99 & 0.989 & 0.98\\
        $t_\mathrm{i}$ & $\sqrt{1-r_\mathrm{i}^2-10^{-5}}$ & $\sqrt{1-r_\mathrm{i}^2-10^{-5}}$ & $\sqrt{1-r_\mathrm{i}^2-10^{-5}}$\\
        $t_\mathrm{e}$ & $\sqrt{1-r_\mathrm{e}^2-10^{-5}}$ & $\sqrt{1-r_\mathrm{e}^2-10^{-5}}$ & $\sqrt{1-r_\mathrm{e}^2-10^{-5}}$\\
        $L$~[m] & $10^5+10.6$ & $10^5+10.2$ & $10^5+10.4$\\
        $\delta f_1$~[Hz] & 25 & 22 & 27\\
        $\delta f_2$~[Hz] & 30 & 27 & 21\\
        \hline
        $a_{00}^2$ & \multicolumn{3}{c}{100~(common for 6 lasers)}\\
        $a_{01}^2$ & \multicolumn{3}{c}{1~(common for 6 lasers)}\\
        $f_\mathrm{m}$ ~[MHz]& \multicolumn{3}{c}{20~(common for 6 lasers)}\\
        \hline
    \end{tabular}
    \caption{Parameters of the three cavities. We assumed a pessimistic case where the values of the parameters make the cavity responses complex so that we can show that our estimation scheme is robust. $r_\mathrm{i, e}$ and $t_\mathrm{i, e}$ represent the amplitude reflectivity and transmittance of the input and end mirror, respectively. $L$ denotes the length of the cavities. The cavity lengths are chosen such that the resonance frequency of the RF sideband is near the resonance frequency of the carrier. This makes the cavity response more difficult to estimate. $\delta f_i$ is the difference of the resonance frequency of the spatial higher order mode from the resonance frequency of the carrier for the laser incident from each side of the cavity. These frequencies are also set to make the response more complex. The ratio of the power of the fundamental Hermite-Gauss 00 mode $a_{00}^2$ is set to be 100 times larger than the higher order $\mathrm{HG}_{01}$ mode $a_{01}^2$.}
    \label{tab:params}
\end{table}
To verify the effectiveness of the proposed null-stream calibration scheme and to evaluate the estimation bias of the cavity responses, we performed numerical simulations using the system parameters listed in Table~\ref{tab:params}. The source code used to generate the simulation is publicly available~\cite{yusuke_okuma_2026_23902816}. To emulate realistic operating conditions, we assumed that both the radio-frequency (RF) sidebands and higher-order spatial modes of the laser fields are coupled to the cavities with finite amplitudes.

The cavity transfer function is modeled as the sum of several response components $S_{\mathrm{com}}(\omega;\delta \omega_i,\omega_b,\omega_c)$.
\begin{eqnarray}
    S_i=\sum_{(\delta \omega_i,\omega_b,\omega_c)} S_{\mathrm{com}}(\omega;\delta \omega_i,\omega_b,\omega_c).
\end{eqnarray}
The derivation of $S_{\mathrm{com}}(\omega;\delta \omega_i,\omega_b,\omega_c)$ is presented in Appendix~\ref{sec:app}. The transfer function $S_i$ depends on nine parameters: the amplitude reflectivities of the input and end mirrors, $r_\mathrm{i}$ and $r_\mathrm{e}$; the corresponding amplitude transmittance, $t_\mathrm{i}$ and $t_\mathrm{e}$; the cavity length $L$; the frequency separation between the higher-order spatial mode and the fundamental mode, $\delta f_i$; the coefficients determining the amplitudes of the fundamental and higher-order-mode responses, $a_{00}$ and $a_{01}$; and the RF modulation frequency $f_\mathrm{m}$.

Since two lasers share the same cavity, several parameters are common to both transfer functions. For the three cavities, the total number of free parameters to be optimized is 39.

The mirror reflectivities were chosen such that the cavity finesse is approximately 100, while the optical loss of each mirror was assumed to be 10~ppm. The cavity lengths were selected to represent a conservative scenario in which the resonance frequencies of the RF sidebands are located close to that of the carrier field. Under this condition, resonance structures appear in the transfer functions, making the calibration problem more challenging. Similarly, the values of $\delta f_i$ were chosen to be close to zero to further increase the complexity of the response functions. The parameters $a_{00}$, $a_{01}$, and $f_\mathrm{m}$ were assumed to be identical for all lasers. In the simulation, the modulation frequency $f_\mathrm{m}$ was fixed to its true value because it is directly determined by the control system. For the simulation, we generated noise in frequency domain with the frequency resolution of 0.001~Hz.

The parameters were divided into three groups Params~$i$, as illustrated in Fig.~\ref{fig:strategy}. Each group corresponds to one cavity and contains the parameters associated with that cavity.

As summarized in Fig.~\ref{fig:strategy}, the process consists of four steps. First, six laser-frequency-noise realizations, six PDH-error-noise realizations, and three beat-noise realizations were generated in the frequency domain as independent Rayleigh random variables. The amplitudes were sampled from Rayleigh distributions with scale parameter
\begin{eqnarray}
    \frac{\Tilde{n}_{\mathrm{b}ij}}{\nu}&=&1.5\times 10^{-23}f~[\mathrm{1/\sqrt{Hz}}]\\
    \Tilde{n}_{\mathrm{e}i}&=&\frac{7}{\sqrt{2}}\times 10^{-23}\frac{S_i(f)}{S_i(0)}~[\mathrm{1/\sqrt{Hz}}]\\
    \frac{\Tilde{\delta \nu}_i}{\nu_i}&=&\frac{10^{-19}}{6}f^{-1}~[\mathrm{1/\sqrt{Hz}}].
\end{eqnarray}
These values follow the proposal paper of the BLFPI~\cite{izumi2021back}.

Next, 20 independent sets of initial parameter values were generated. Each initial value was obtained by adding Gaussian noise with zero mean and a fractional standard deviation of 1\% to the corresponding true parameter value. Optimization was then performed starting from each of the twenty initial parameter sets.

Using these noise realizations and parameter values, we evaluated the cost function defined as the IP of the null channel [Eq.~(\ref{eq:iir:null})]. To prevent the cost function from being dominated by high-frequency components, a frequency-dependent weighting function was introduced. We adopted

\begin{eqnarray}
    w(\omega)=
    \left\{\begin{array}{ll}
        1 & :\omega<2\pi\times 0.1~[\mathrm{Hz}] \\
        0 & :2\pi\times 0.1~[\mathrm{Hz}]<\omega
    \end{array}\right.\label{eq:weight}
\end{eqnarray}

Because the optimization algorithm cannot reliably handle cost function values below $10^{-15}$, the cost function was divided by a value N. We used $N=5\times 10^{-43}$ The resulting cost function $C$ is

\begin{eqnarray}
    &&C\notag\\
    &=&\frac{1}{N}\int_{0}^\infty d\omega~w(\omega) \left\{\sum^6_{i=1}\left|\left(1-\frac{S_i}{\hat{S}_i}\right)\frac{1}{1+G_i}\right|^2\frac{|\Tilde{\delta\nu}_i(\omega)|^2}{\nu^2}\notag\right.\\
    &&\left.+\sum^6_{i=1}\left|\frac{\Tilde{n_{\mathrm{e}i}}(\omega)}{\hat{S}_i}\right|^2+\sum_{i=1,3,5}|\Tilde{n}_{\mathrm{b}i,i+1}(\omega)|^2\right\}\label{eq:iir:cost}.
\end{eqnarray}
The parameters were optimized so as to minimize this cost function.

To reduce the dimensionality of the optimization problem, only one parameter group was optimized at a time while all other groups were fixed. Parameter optimization was performed using the $\texttt{least\_squares}$ provided by $\texttt{SciPy}$~\cite{scipy_least_squares}. For each optimization step, the maximum number of function evaluations ("maxnfev") was set to 10.

The optimization cycle over the three parameter groups was repeated twenty times. After each cycle, the resulting cost value was compared with the threshold of $5\times10^{41}$. If the cost remained above the threshold, the optimization was terminated and restarted from a newly generated initial parameter set. Otherwise, the optimization was continued until no further reduction in the cost function was observed. This threshold was introduced to avoid excessive computational resources being spent on optimization runs that were unlikely to converge to the global minimum.

After all twenty initial parameter sets had been optimized, the solution yielding the lowest cost value was selected. In total, 1000 independent realizations were evaluated, with new noise realizations generated for each run while maintaining the true parameter values unchanged.

\begin{figure*}[!t]
    \centering
    \includegraphics[width=1\linewidth]{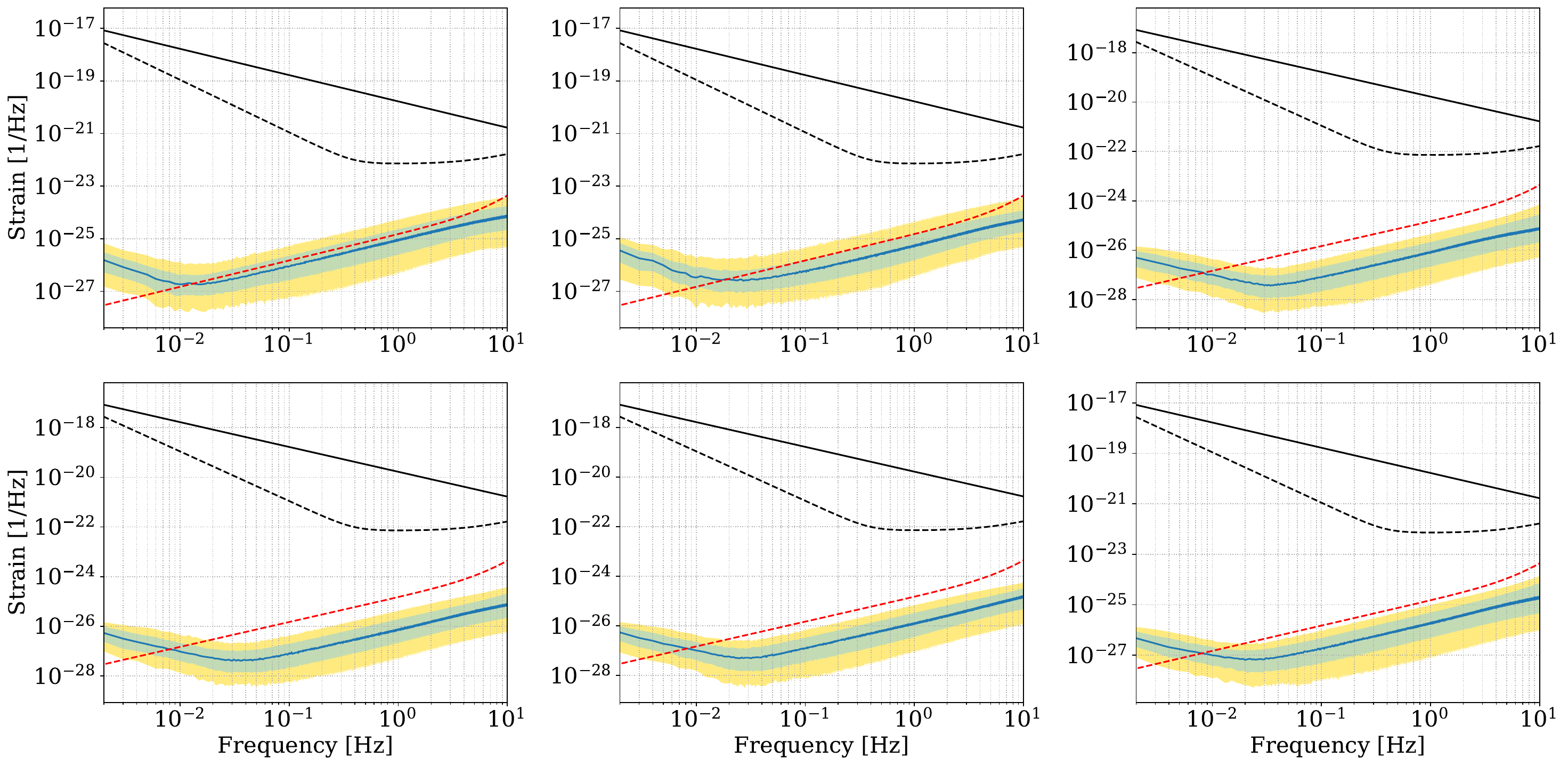}
    \caption{Residual frequency noise for the six lasers. The panels in the first row, from left to right, correspond to Lasers 1, 2, and 3, respectively, and those in the second row, from left to right, correspond to Lasers 4, 5, and 6, respectively. Lasers 1 and 2 are incident on Cavity 1, Lasers 3 and 4 on Cavity 2, and Lasers 5 and 6 on Cavity 3. The cavity parameters are given in Table~\ref{tab:params}. The shaded regions represent the 95\% and 68\% occurrence intervals at each frequency. The blue line represents the median value. The dashed black line shows the proposed BLFPI sensitivity in~\cite{izumi2021back}. The solid black line shows the expected frequency noise. The dashed red line represents the typical residual frequency noise using Eq.~(\ref{eq:iir:one_res}), (\ref{eq:iir:biased}) and (\ref{eq:iir:biasp}).}
    \label{fig:res_noise}
\end{figure*}

Figure~\ref{fig:res_noise} shows the residual frequency noise for the six lasers in the units of strain. The panels in the first row, from left to right, correspond to lasers 1, 2, and 3, respectively, and those in the second row, from left to right, correspond to lasers 4, 5, and 6, respectively. Lasers 1 and 2 are incident on Cavity 1, Lasers 3 and 4 on Cavity 2, and Lasers 5 and 6 on Cavity 3. The shaded regions indicate the 68\% and 95\% occurrence intervals obtained from 1000 independent realizations.

For all six lasers, the 95\% occurrence intervals remain more than an order of magnitude below the designed BLFPI sensitivity throughout the frequency range below 10~Hz. Between 0.02 and 2~Hz, the median residual frequency noise is several times lower than the analytically predicted typical level derived from the single frequency component analysis (Eq.~(\ref{eq:iir:opt_noise})). This discrepancy arises from the setup of the simulation, in which six lasers are resonant in three cavities. In this configuration, the estimations of the cavity responses for the six lasers are not independent, and their estimation biases are correlated. This differs from the setup assumed in the analytical derivation, in which one laser is resonant in one cavity. Below 0.02~Hz, the residual noise increases. We confirmed that this increase originates from numerical errors in the evaluation of the response function given by Eq.~(\ref{eq:iir:pdh}). Above 3~Hz, the median residual noise decrease further relative to the analytically derived typical level. This is because the numerical simulation utilize all the frequency components simultaneously rather than single frequency component. Consequently, the typical residual noise predicted analytically is not necessarily realized.

\section{Discussion\label{sec:discussion}}
As demonstrated in the numerical simulations, the design of the cost function of the proposed calibration method is open. In Sec.~\ref{sec:num}, we adopted the simple weighting function given by Eq.~(\ref{eq:weight}). If the IP of the null channel is evaluated without any weighting, high-frequency components dominate the cost function because they contribute more strongly to the IP. Consequently, the optimization tends to prioritize fitting the cavity response at high frequencies at the expense of the lower-frequency region that is most relevant for GW observations. Therefore, the weighting function should be refined for achieving high calibration accuracy in the observation band. For practical implementation in a space-based detector, the weighting function should be optimized by taking into account the expected spectra of laser frequency noise, technical noise, and the target detector sensitivity. Determining such an optimal weighting scheme remains an important subject for future work.

Another important issue is the time variation of the cavity response. In a space-based interferometer, the distances between spacecraft vary continuously because of orbital motion. As a result, the cavity lengths and their transfer functions also change with time. The numerical simulations presented in this work assumed static cavity responses over a period of 1000~s and therefore did not account for such temporal variations. To apply the proposed calibration method to an actual space mission, it will be necessary to estimate the cavity responses continuously and track their time evolution. This will likely require the development of adaptive calibration algorithms capable of updating the response functions in real time~\cite{tuyenbayev2017improving}. Evaluating the performance of the proposed null-stream calibration method under realistic time-varying conditions will be an important topic for future investigation.

An important advantage of the proposed calibration method is that it has the potential to achieve absolute calibration of the cavity responses with relatively small systematic uncertainty. In the BLFPI, the GW signal is reconstructed from the beat frequency between two independent lasers. Because the response of the beat-frequency readout system, $K$, is directly connected to the on board clock, its response can be accurately calibrated before launch and its systematic uncertainty is expected to be much smaller than that associated with the cavity responses. Consequently, the proposed method provides an absolute calibration of the cavity responses rather than relying on relative calibration between different interferometers.

This feature distinguishes the proposed method from null-stream calibration techniques developed for networks of ground-based GW detectors~\cite{schutz_self-calibration_2020,gursel_near_1989}, where the detector responses can only be calibrated relative to one another. Absolute calibration is particularly important for future space-based missions targeting a stochastic or primordial GW background. To extract a stochastic background GW, foreground GW signals must be subtracted using the detector response. Any systematic error in the detector response can leave residual foreground signals, resulting in an error in the inferred GW energy density. This capability may therefore become one of the key advantages of the BLFPI for future precision GW observations. In searches for a primordial GW background, the target GW energy density is expected to be extremely small. Consequently, calibration errors in the detector response directly limit the achievable measurement accuracy and may become one of the dominant sources of systematic uncertainty. The proposed calibration method therefore provides a promising approach for reducing such systematic errors and enabling future high-precision measurements of stochastic and primordial GW backgrounds.

\section{Conclusion}\label{sec:conclusion}
In this paper, we proposed a novel calibration scheme for estimating the cavity response functions in the Back-linked Fabry--Perot Interferometer (BLFPI). The proposed method utilizes a null channel to suppress the influence of GW signals, which would otherwise hinder accurate calibration. By optimizing the cavity-response parameters to minimize the integrated power value of the null channel, the cavity responses can be estimated without requiring additional calibration hardware.

We analytically investigated the accuracy limit of the proposed method and showed that technical noise in the PDH error signals introduces a bias in the estimated cavity responses. We confirmed the expected bias was found to be sufficiently small and is not expected to significantly degrade the frequency-noise suppression performance of the BLFPI.

To evaluate the feasibility of the proposed scheme, we performed numerical simulations using realistic cavity models that included contributions from RF sidebands and higher-order spatial modes. The simulation results demonstrated that the residual frequency noise after calibration remains more than an order of magnitude below the designed BLFPI sensitivity throughout the observation band. These results indicate that the proposed null-channel calibration method can achieve the accuracy required for effective subtraction of laser frequency noise.

Several topics remain for future investigation. In particular, the design of a weighting function used in the cost function should be optimized to maximize calibration accuracy in the observation band. In addition, practical implementation in a space-based detector will require consideration of the time-varying cavity responses caused by changes in the spacecraft separation. Extending the proposed method to account for such temporal variations will be an important step toward its application in future space-based GW observatories.

Overall, the proposed null-channel calibration method provides a practical strategy for accurately estimating cavity response functions without dedicated calibration hardware. This capability has the potential to reduce systematic calibration errors in future space-based GW detectors and support the high-precision measurements required for observations of stochastic and primordial GW backgrounds.

\section{Acknowledgment}
This work was supported by JSPS KAKENHI Grant Number JP24K00651. We thank S. Tada, S. Miyazaki and S. Shibai for fruitful discussions.

\vspace{0.2cm}
\noindent
\bibliographystyle{apsrev.bst}
\bibliography{paper}

\section{Appendix}
\subsection{Derivation of the cavity response function\label{sec:app}}
The designed finesse of the arm cavities of the BLFPI is 100. Under this condition, RF sidebands and higher-order spatial modes can become resonant within the cavities, resulting in a complicated frequency-dependent response. Here, we derive the cavity response while taking into account the resonance of the RF sidebands and spatial higher-order modes.

Since we measure the frequency noise with PDH method, we should consider nine angular frequency components below, namely, $\Omega$, $\Omega\pm \omega_a$, $\Omega\pm\omega_\mathrm{m}$ and $\Omega\pm\omega_\mathrm{m}\pm \omega_a$, where $\Omega/2\pi$, $\omega_a/2\pi$ and $\omega_\mathrm{m}/2\pi$ are the carrier-laser frequency, frequency-noise frequency and RF modulation frequency applied to the EOM, respectively. Each of the nine frequency components is reflected from the cavity. The complex amplitude of the reflected laser field, $E_\mathrm{r}$ is written as
\begin{eqnarray}
    E_\mathrm{r}=\sum_{\omega=\Omega,~\Omega\pm \omega_a,~\Omega\pm\omega_\mathrm{m},~\Omega\pm\omega_a\pm\omega_\mathrm{m}}r(\omega)E_{\mathrm{in}}(\omega),
\end{eqnarray}
where $r(\omega)$ is the complex reflectivity of the cavity. It is given by
\begin{eqnarray}
    r(\omega)
    &=&\left(-r_{\mathrm{i}}+\frac{r_{\mathrm{e}}t_{\mathrm{i}}e^{-2i\omega T_\mathrm{c}}}{1-r_{\mathrm{i}}r_{\mathrm{e}}e^{-2i\omega T_\mathrm{c}}}\right).\label{eq:iir:cav_reflectivity}
\end{eqnarray}
The PDH signal $V_{\mathrm{PDH}}$ is obtained by demodulating the power of the reflected laser field:
\begin{eqnarray}
    V_{\mathrm{PDH}}=\left|E_\mathrm{r}\right|^2\sin{\omega_\mathrm{m}t}.
\end{eqnarray}
To calculate the response of the cavity to frequency noise, we only need to consider the terms proportional to $e^{\pm i\omega_a t}$. The component of the PDH signal proportional to the frequency noise is therefore given by a combination of the cavity reflectivities:
\begin{eqnarray}
    &&V_{\mathrm{PDH}}\notag\\
    &\propto&\Re\left[\{-r(\Omega-\omega_a)^*r(\Omega+\omega_\mathrm{m})-r(\Omega+\omega_a)r(\Omega-\omega_\mathrm{m})^*\right.\notag\\
    &&-r(\Omega+\omega_a)r(\Omega+\omega_\mathrm{m})^*-r(\Omega-\omega_a)^*r(\Omega-\omega_\mathrm{m})\notag\\
    &&r(\Omega)^*r(\Omega+\omega_\mathrm{m}+\omega_a)+r(\Omega)r(\Omega-\omega_\mathrm{m}-\omega_a)^*\notag\\
    &&\left.r(\Omega)r(\Omega+\omega_\mathrm{m}-\omega_a)^*+r(\Omega)^*r(\Omega-\omega_\mathrm{m}+\omega_a)\}e^{i\omega_a t}\right].\notag\\
    &&\label{eq:iir:pdh_sb}
\end{eqnarray}
 There are two types of interference. The first is the interference between the frequency-modulated sidebands, with frequencies $(\Omega\pm\omega_\mathrm{a})/2\pi$, and the RF sidebands, with frequencies $(\Omega\pm\omega_\mathrm{m})/2\pi$. The second is the interference between the carrier field, with frequency $\Omega/2\pi$ and the RF sidebands carrying frequency noise with frequencies $(\Omega\pm\omega_\mathrm{m}\pm\omega_\mathrm{a})/2\pi$. The cavity response to frequency noise arising from these two types of interference are given by
\begin{eqnarray}
    &&S_{00}\notag\\
    &=&a_{00}^2\left\{-r(\Omega-\omega_a)^*r(\Omega+\omega_\mathrm{m})-r(\Omega+\omega_a)r(\Omega-\omega_\mathrm{m})^*\right.\notag\\
    &&\left.-r(\Omega+\omega_a)r(\Omega+\omega_\mathrm{m})^*-r(\Omega-\omega_a)^*r(\Omega-\omega_\mathrm{m})\right\}\notag\\
    &&+a_{00}^2\left\{r(\Omega)^*r(\Omega+\omega_\mathrm{m}+\omega_a)+r(\Omega)r(\Omega-\omega_\mathrm{m}-\omega_a)^*\right.\notag\\
    &&\left.r(\Omega)r(\Omega+\omega_\mathrm{m}-\omega_a)^*+r(\Omega)^*r(\Omega-\omega_\mathrm{m}+\omega_a)\right\},\notag\\
    &&
\end{eqnarray}
where $a_{00}^2$ represents the product of the amplitude of the carrier field and RF sideband, the demodulation efficiency, and the response of the PD. The first term represents the interference between the frequency-modulated sidebands, with frequencies $(\Omega\pm\omega_\mathrm{a})/2\pi$, and the RF sidebands, with frequencies $(\Omega\pm\omega_\mathrm{m})/2\pi$. The second term represents the interference between the carrier field, with frequency $\Omega/2\pi$ and the RF sidebands carrying frequency noise with frequencies $(\Omega\pm\omega_\mathrm{m}\pm\omega_\mathrm{a})/2\pi$.

For higher order spatial modes, their carrier fields are generally off resonance, and their resonance frequencies are shifted from that of the fundamental mode by $\delta \omega/2\pi$. By replacing $\Omega$ with $\Omega+\delta\omega$ in Eq.~(\ref{eq:iir:pdh_sb}), the cavity response for a higher order spatial mode can be derived in the same manner as that for the fundamental mode.

\begin{figure}[t]
    \centering
    \includegraphics[width=1\linewidth]{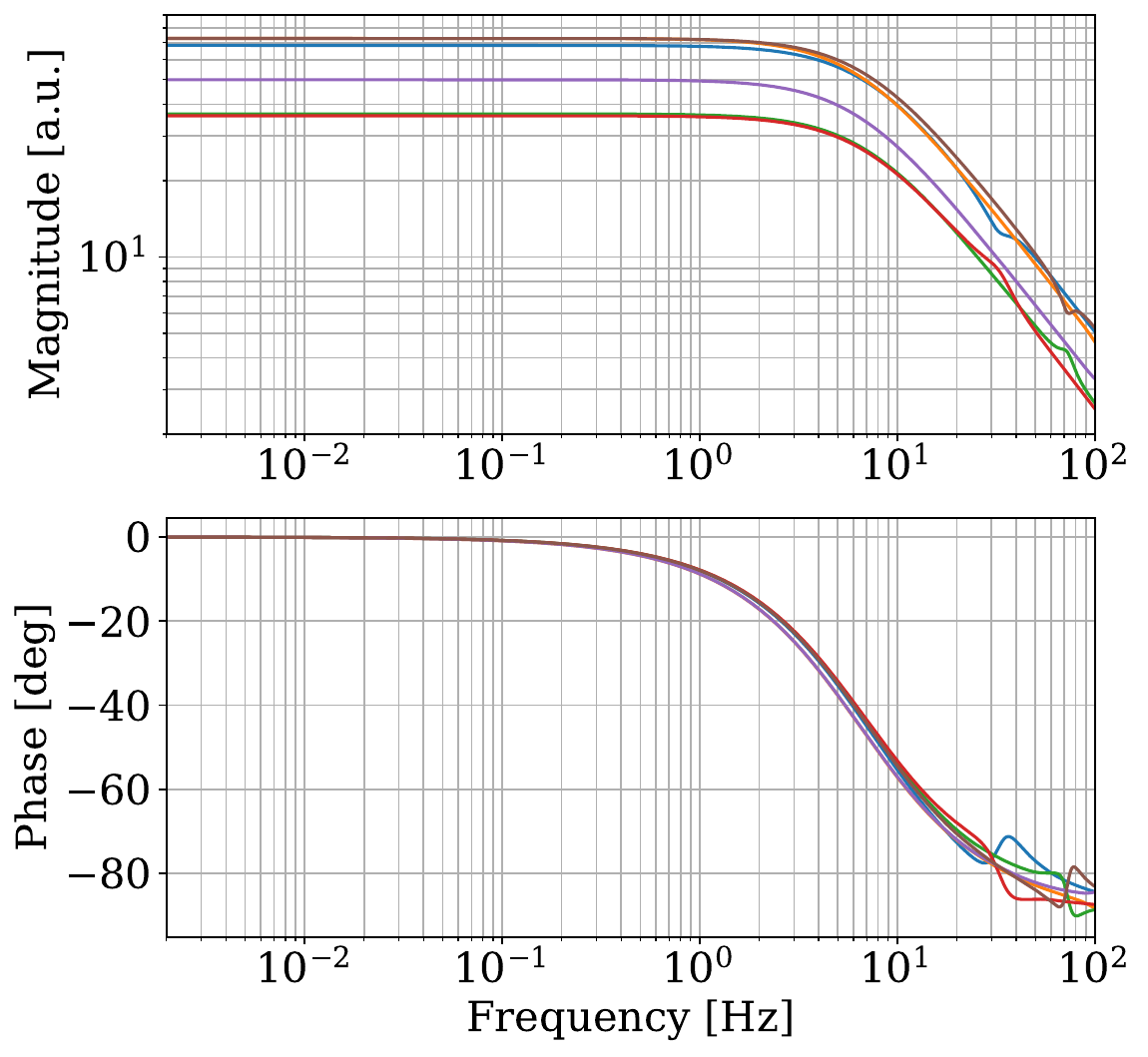}
    \caption{Six cavity responses adopted in the simulation. These responses are calculated with the parameters in the table~\ref{tab:params}.}
    \label{fig:app:pdh_func}
\end{figure}

By expanding the cavity response using Eq.~(\ref{eq:iir:cav_reflectivity}), for all interferences including that of the higher order modes, the PDH response can be expressed in a combination of a unified form:
\begin{eqnarray}
    &&S_\mathrm{com}(\omega_\mathrm{a}; \delta\omega, \omega_b, \omega_c)\notag\\
    &=&a_{ij}^2\left[-4r_{\mathrm{i}}^2\Re\left(1-\frac{r_\mathrm{e}}{r_\mathrm{i}}t_\mathrm{i}^2\gamma\right)+\frac{4r_{\mathrm{i}}r_{\mathrm{e}}t_{\mathrm{i}}^2e^{-2i\omega_aT_{\mathrm{c}}}}{1-r_{\mathrm{i}}r_{\mathrm{e}}e^{2i(\delta\omega+\omega_c-\omega_a)T_{\mathrm{c}}}}\right.\notag\\
    &\times& \left[\left.\frac{\Re\left\{e^{2i(\delta\omega+\omega_c) T_{\mathrm{c}}}\left(1-\frac{r_\mathrm{e}}{r_\mathrm{i}}t_\mathrm{i}^2\gamma\right)\right\}}{1-r_{\mathrm{i}}r_{\mathrm{e}}e^{-2i(\delta\omega+\omega_c+\omega_a)T_{\mathrm{c}}}}\right.\right.\notag\\
    &-&\left.\left.\frac{r_\mathrm{i}r_\mathrm{e}\Re\left(1-\frac{r_\mathrm{e}}{r_\mathrm{i}}t_\mathrm{i}^2\gamma\right)e^{-2i\omega_a T_{\mathrm{c}}}}{1-r_{\mathrm{i}}r_{\mathrm{e}}e^{-2i(\delta\omega+\omega_c+\omega_a)T_{\mathrm{c}}}}\right]\right],\notag\\
    &&\label{eq:iir:pdh}
\end{eqnarray}
where
\begin{eqnarray}
    &&\gamma=\frac{(\cos{2\omega_bT_\mathrm{c}}-r_\mathrm{i}r_\mathrm{e}e^{-2i\delta\omega T_{\mathrm{c}}})e^{-2i\delta\omega T_{\mathrm{c}}}}{\{1-r_{\mathrm{i}}r_{\mathrm{e}}e^{-2i(\delta\omega+\omega_b)T_{\mathrm{c}}}\}\{1-r_{\mathrm{i}}r_{\mathrm{e}}e^{-2i(\delta\omega-\omega_b)T_{\mathrm{c}}}\}}.\notag\\
    &&
\end{eqnarray}
The parameters $\delta\omega$, $\omega_b$ and $\omega_c$ are assigned the values listed in Table~\ref{tab:iir:hom_param} for each interference term.
\begin{table}[b]
    \centering
    \begin{tabular}{c|c|c|c|c|c}
         & Interference & Ratio & $\delta\omega$ & $\omega_b$ & $\omega_c$ \\
         \hline
        $\mathrm{HG}_{00}$ & $\mathrm{dC*RF}$ & 1 & 0 & $\omega_\mathrm{m}$ & $0$ \\
        & $\mathrm{C*dRF}$ & -1 & $0$ & $0$ & $\omega_\mathrm{m}$\\
        \hline
        $\mathrm{HG}_{ij}$ & $\mathrm{dC*RF}$ & 1 & $\delta \omega$ & $\omega_\mathrm{m}$ & $0$\\
        & $\mathrm{C*dRF}$ & -1/2 & $\delta \omega$ & 0 & $\omega_\mathrm{m}$\\
        & $\mathrm{C*dRF}$ & -1/2 & $\delta \omega$ & 0 & $-\omega_\mathrm{m}$
    \end{tabular}
    \caption{Parameters used in the cavity response calculation. The "Interference" column indicates the frequency components involved in each interference term. $\mathrm{dC*RF}$ denote the interference between the carrier field with frequency fluctuations and the RF sidebands without frequency fluctuations. $\mathrm{C*dRF}$ denotes the interference between the carrier field without frequency fluctuations and the RF sidebands with frequency fluctuation. The "Ratio" column gives the ratio of the response, represented by $a_{ij}^2$, to the $\mathrm{dC*RF}$ response of each spatial mode.}
    \label{tab:iir:hom_param}
\end{table}
By summing all interference terms, the cavity response to laser frequency fluctuations is given by
\begin{eqnarray}
    S_i(\omega_\mathrm{a})=\sum_{( \delta\omega, \omega_b, \omega_c)} S_{\mathrm{com}}(\omega_\mathrm{a}; \delta\omega, \omega_b, \omega_c).
\end{eqnarray}
Using this formula with the parameters in the table~\ref{tab:params}, the cavity responses are calculated and shown in Fig.~(\ref{fig:app:pdh_func}).

\subsection{Noise level of the BLFPI after optimization\label{sec:app:noise}}
This appendix explains why optimizing the cavity response appears to reduce the residual noise, even though the detector sensitivity is fundamentally limited by shot noise. We show that the apparent noise reduction reflects the partial recovery of the signal-to-noise ratio, which is degraded by the beam splitting that divides the reflected light between the in-loop and out-of-loop PDs, and therefore does not violate the shot-noise limit.

\begin{figure}[t]
    \centering
    \includegraphics[width=1\linewidth]{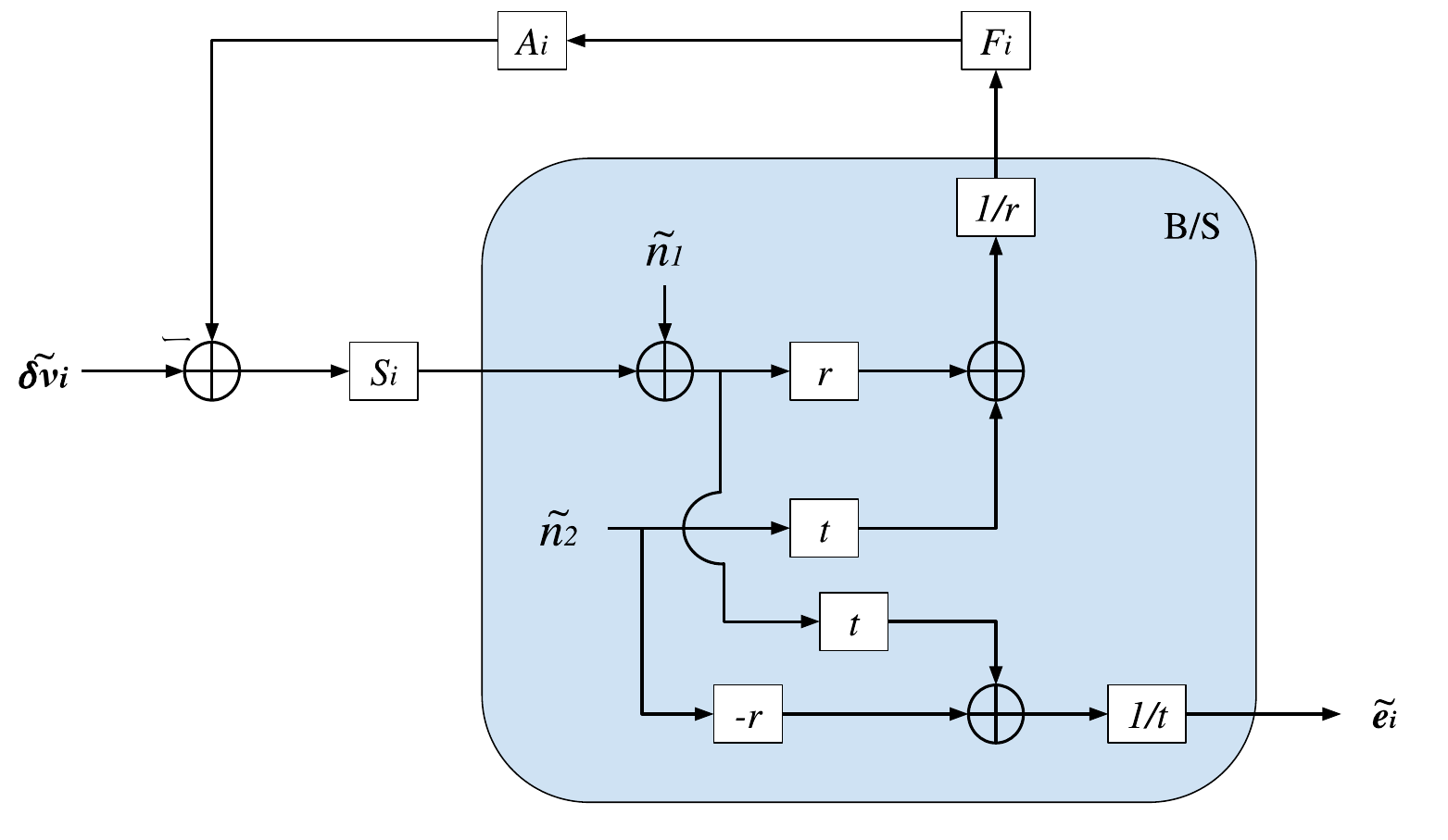}
    \caption{Block diagram of the laser-frequency control system for laser $i$. The colored rounded rectangle corresponds to the region enclosed by the dashed line in Fig.~\ref{fig:block}. Vacuum fluctuations $\Tilde{n}_1$ and $\Tilde{n}_2$ enter through the beam splitter that divides the cavity-reflected beam between the in-loop and out-of-loop photo detectors.}
    \label{fig:null:vacuum_noise}
\end{figure}
Eq.~(\ref{eq:null:noise_opt}) indicates that optimizing the cavity response slightly reduces the residual noise of the BLFPI. At first glance, this result appears counterintuitive because the noise floor is ultimately limited by shot noise and it is impossible to decrease shot noise without the use of higher optical power or quantum resources.

To understand the origin of this behavior, we consider the vacuum fluctuations entering through the open port of the beam splitter. Fig~\ref{fig:null:vacuum_noise} shows the block diagram of the laser frequency control system for laser $i$. Here, $\Tilde{n}_1$ represents the intrinsic vacuum fluctuation coming from the arm cavity while $\Tilde{n}_2$ is the vacuum fluctuation entering from the open port of the beam splitter. $\Tilde{n}_1$ and $\Tilde{n}_2$ are independent. The beat signal and the PDH error signal are then given by
\begin{eqnarray}
    \Tilde{b}_i
    &=&\frac{1}{1+G_i}\Tilde{\delta \nu}_i-\frac{1}{S_i}\frac{G_i}{1+G_i}\Tilde{n}_1-\frac{1}{S_i}\frac{G_i}{1+G_i}\frac{t}{r}\Tilde{n}_2\label{eq:iir:beat_noise}\\
    \Tilde{e}_i
    &=&\frac{S_i}{1+G_i}\Tilde{\delta \nu}_i+\frac{1}{1+G_i}\Tilde{n}_1-\frac{G_i}{1+G_i}\frac{t}{r}\Tilde{n}_2-\frac{r}{t}\Tilde{n}_2.\notag\\
    &&\label{eq:iir:error_noise}
\end{eqnarray}
The effective shot-noise contributions to the in-loop and out-of-loop photo detectors are
\begin{eqnarray}
    \Tilde{n'}&=&\Tilde{n}_1+\frac{t}{r}\Tilde{n}_2\\
    \Tilde{n}&=&\Tilde{n}_1-\frac{r}{t}\Tilde{n}_2,
\end{eqnarray}
respectively. Note that $\Tilde{n'}$ and $\Tilde{n}$ are uncorrelated. Using $\Tilde{n'}$ and $\Tilde{n}$, Eqs.~(\ref{eq:iir:beat_noise}) and (\ref{eq:iir:error_noise}) can be written as
\begin{eqnarray}
    \Tilde{b}_i&=&\frac{1}{1+G_i}\Tilde{\delta \nu}_i-\frac{1}{S_i}\frac{G_i}{1+G_i}\Tilde{n'}\\
    \Tilde{e}_i&=&\frac{S_i}{1+G_i}\Tilde{\delta \nu}_i-\frac{G_i}{1+G_i}\Tilde{n'}+\Tilde{n}.
\end{eqnarray}
After post process considering only laser $i$, the BLFPI signal is given by
\begin{eqnarray}
    &&\Tilde{b}_i-\frac{\Tilde{e}_i}{\hat{S_i}}\\
    &=&\frac{1}{1+G_i}\Tilde{\delta \nu}_i\left(1-\frac{S_i}{\hat{S}_i}\right)-\frac{1}{S_i}\frac{G_i}{1+G_i}\Tilde{n'}\left(1-\frac{S_i}{\hat{S}_i}\right)-\frac{\Tilde{n}}{\hat{S}_i}.\notag\\
    &&
\end{eqnarray}
The optimization minimizes the IP of this quantity. Assuming that the two vacuum fluctuations have equal power, $|\Tilde{n}_1|^2=|\Tilde{n}_2|^2$, the power spectrum can be written as
\begin{eqnarray}
    &&\left|\Tilde{b}_i-\frac{\Tilde{e}_i}{\hat{S_i}}\right|^2\\
    &=&\left\{\left|\frac{\Tilde{\delta\nu}_i}{1+G_i}\right|^2+\left|\frac{G_i}{1+G_i}\right|^2\frac{|\Tilde{n}_1|^2}{R|S_i|^2}\right\}\left|1-\frac{S_i}{\hat{S}_i}\right|^2\notag\\
    &&+\frac{1}{T|\hat{S}_i|^2}|\Tilde{n}_1|^2\\
    &\geq&\left|\frac{\Tilde{n}_1}{S_i}\right|^2\frac{1}{T+\frac{1}{\left|\frac{S_i}{\Tilde{n}_1}\right|^2\left|\frac{\Tilde{\delta\nu}_i}{1+G_i}\right|^2+\frac{1}{R}\left|\frac{G_i}{1+G_i}\right|^2}}\\
    &\geq&\left|\frac{\Tilde{n}_1}{S_i}\right|^2\label{eq:iir:RMS_min}
\end{eqnarray}
where $R$ and $T$ denote the power reflectivity and transmittance of the beam splitter, respectively. Eq.~(\ref{eq:iir:RMS_min}) shows that the optimization cannot reduce the noise below the shot noise level determined by the intrinsic vacuum fluctuation $\Tilde{n}_1$. In the limit of infinite loop gain, the residual noise asymptotically approaches the shot-noise limit:
\begin{eqnarray}
    \left|\Tilde{b}_i-\frac{\Tilde{e}_i}{\hat{S_i}}\right|^2
    &\geq&\left|\frac{\Tilde{n}_1}{S_i}\right|^2\frac{1}{T+\frac{1}{\left|\frac{S_i}{\Tilde{n}_1}\right|^2\left|\frac{\Tilde{\delta\nu}_i}{1+G_i}\right|^2+\frac{1}{R}\left|\frac{G_i}{1+G_i}\right|^2}}\\
    &\rightarrow&\left|\frac{\Tilde{n}_1}{S_i}\right|^2~(G_i\rightarrow \infty)
\end{eqnarray}

Therefore, the apparent reduction in the residual noise reflects the partial recovery of the signal-to-noise ratio, which is degraded by the beam splitting between the in-loop and out-of-loop PDs, rather than from any reduction of the fundamental shot-noise limit itself.

\end{document}